# Improved Median Polish Kriging for Simulation Metamodeling

Firas Al Rekabi [1] and Asim El Sheikh [2]

[1] Irbid National University, Irbid 2600, Jordan

Firasajil@yahoo.com

[2] University of Banking and Financial Sciences, Amman 11942, Jordan
A.Elsheikh@aabfs.org

**Abstract** In simulation, Median Polish Kriging is a technique used to predict unobserved data points in two-dimensional space. The linear behavior of the traditional Median Polish Kriging in the estimation of the mean function in a high grid makes the interpolation of O(1) which has a low order in the prediction and that leads to a high prediction error. Therefore, an improvement in the estimation of the mean function has been introduced using Biharmonic spline interpolation and the new technique has been called Improved Median Polish Kriging (IMPK). The IMPK has been applied to the standard coal-ash data in two-dimension. The novel method gave much better results according to the cross validation results that were obtained when compared with the traditional Median Polish Kriging.

**Keywords:** Simulation, Metamodels, Kriging, Median Polish

## 1. Introduction

In some simulation experiments, we might want to use the observed data themselves to specify directly (in some sense) a distribution, this is called an empirical distribution, from which random values are generated during the simulation, rather than fitting a theoretical distribution to the data. Simulation models are often tedious to build, need substantial data for input modeling, and require significant time to run, particularly when there are many alternatives to evaluate. Consequently, statistical approximations are becoming widely used in engineering to construct simplified approximations, or metamodels, of these analysis codes that are then used instead of the actual analysis codes, providing a surrogate model of the original code. A metamodel is an approximation of the input/output (I/O) transformation that is implied by the simulation model; the resulting black-box model is also known as a response surface or emulator [4]. The most widely used method for metamodels in simulation is Kriging. Kriging is a group of geostatistical techniques to interpolate the value of a random field at an unobserved location from observations of its value at nearby locations. The first use of kriging in simulation was found by [11].

Kriging technique was originally evolved in geostatistics by D. G. Krige [5], and has recently been widely applied in deterministic simulation, it gives more weight to 'neighboring' observations. Actually, Kriging give quite acceptable predictions; traditional linear regression gives the worst results [10]. Kriging provides exact interpolation, i.e., the predicted output values at 'old' input combinations already observed are equal to the simulated output values at those inputs ('inputs' are also called 'factors'; 'input combinations' are also called 'scenarios') [3]. Obviously, such interpolation is appealing in deterministic simulation. Kriging and deterministic simulation are often applied in Computer Aided Engineering (CAE) for the (optimal) design of airplanes, automobiles, computer chips, computer monitors, etc.; [6].

## 2. Kriging Preliminaries

The primary motivation behind the use of Kriging in most earth science applications, and one of the essential reasons for its introduction, is that it is non-parametric. Moreover, the kriging model has been used as a metamodel in the design and analysis of computer experiments (DACE) [6]. In the application of kriging model in the field of simulation, the parameters of the model are likely to be estimated from the simulated data. In building the kriging model and its predictor, in addition to the sample observations, the best linear unbiased estimator (BLUE) depends on the parameters in the mean and the covariance parameters. In an ideal situation, these parameters are assumed known. In practice however, these parameters can only be estimated from sample data, making them random variables dependent on the experimental design and sample observations.

Mathematically speaking, a random process Z(.) can be described by {Z(s):s∈D} where D is a fixed subset of Rd and Z(s) is a random function at locations $s_1, s_2, \ldots, s_n$; [2]. The basic form of the kriging estimator is,

$$Z^*(s) = \sum_{i=1}^{n} \lambda_i Z(s_i) \qquad (1)$$

Where $Z(s_1), Z(s_2), \ldots, Z(s_n)$ are observed values which are obtained at the nth known locations $s_1, s_2, \ldots, s_n$ in solution space, which shows an estimated value of $Z^*(s)$ at $s \in D$, which is the point where we want to estimate the value of the function, also we may note that:

$$\sum_{i=1}^{n} \lambda_i = 1 \qquad (2)$$

Actually, the goal is to determine weights, $\lambda_i$'s that minimize the variance of the estimator, [11], is:

$$\sigma_E^2(s) = Var\{Z^*(s) - Z(s)\}$$

under the unbiasedness constraint

$$E\{Z^*(s) - Z(s)\} = 0$$

Now, we assume that the trend component is a constant and known mean, m (s) = m, so that

$$Z^*(s) = m(s) + \sum_{i=1}^{n} \lambda_i [Z(s_i) - m(s_i)]$$

This estimate is automatically unbiased, so that E[Z*(u)]=m=E[Z(u)]. The estimation error Z*(u)-Z(u) is a linear combination of random variables representing residuals at the data points, $s_i$, and the estimation point, s, [9]:

$$Z^*(s) - Z(s) = E[Z^*(s) - m] - [Z(s) - m] \quad (3)$$

We can write (3) in another form as,

$$\sum_{i=1}^{n} \lambda_i(s) R(s_i) - R(s) = R^*(s) - R(s)$$

Using rules for the variance of a linear combination of random variables, the error variance is then given by

$$\sigma_E^2(s) = Var\{R^*(s)\} + Var\{R(s)\} - 2Cov\{R^*(s), R(s)\}$$
$$= \sum_{i=1}^{n}\sum_{j=1}^{n} \lambda_i(s)\lambda_j(s) C_R(s_i - s_j) + C_R(0) - 2\sum_{i=1}^{n} \lambda_i(s) C_R(s_i - s)$$

To minimize the error variance, we take the derivative of the above expression with respect to each weight of the kriging weights and set each derivative to zero. This leads to the following system of equations:

$$\sum_{j=1}^{n} \lambda_j(s) C_R(s_i - s_j) = C_R(s_i - s)$$

Since the mean is assumed to be constant in ordinary kriging, the covariance function for Z(u) is the same as that for the residual component, $C(h)=C_R(h)$, so that we can write the simple kriging system directly in terms of C(h):

$$\sum_{j=1}^{n} \lambda_j(s) C(s_i - s_j) = C(s_i - s)$$

As a result, this can be written as system of simultaneous equations in matrix form as [7],

$$K\lambda(u) = k$$

where K is the matrix of covariance between data points, with elements $K_{i,j}=C(s_i-s_j)$, k is the vector of covariance between the data points and the estimation point, with elements given by $k_i = C(s_i-s)$, and $\lambda(s)$ is the vector of simple kriging weights. After that, we can solve for the above system for kriging weights as:

$$\lambda = K^{-1} k$$

Finally, it should be observed that stationary of the variogram is not a necessary requirement for kriging; it is assumed for pragmatic reasons, to allow the variogram to be estimated from the data, [2].

**3. Median Polish Kriging**

Median Polish Kriging (MPK) was introduced by [2], it is a hybrid method combining both Kriging and linear spline interpolation to predict a two-dimensional surface for spatial data. The median polish algorithm gives as an estimate of the mean component as

data = all effect + row effect + column effect + residual by subtracting the medians of each row from the row values, then the medians of the columns from the column values, and recording them in the row effect and column effect variables. This process is repeated until convergence, that is, until the row and column medians are 0.

Spatial data can be thought of Median Polish as a partial sampling of a realization of a random process { Z(s) : s∈D}, and may be represented by the following formula:

$$Z(s) = \mu(s) + R(s) \quad (4)$$

Now, μ(.) is the mean structure and R(.) is the residual structure. In reality, μ(.) is not known, in dimensions higher than one, it is natural to assume μ(.) decomposes additively into directional components, [2]. In this article, our concern is in $R^2$. Therefore, assume:

μ(s) = a + r( x ) + c( y )

Where $a$ is the overall effect using Median Polish and r(x) is the row effect and c(x) is the column effect. Furthermore, the points {$s_i$: i=1,2,…n} are actually on a grid { $(x_l,y_k)$ : l=1,2,…q, k=1,2,…p}.

Now, μ(.) given in equation (4) satisfying the values at the grid point only. Hence, to interpolate the data between the exact grid points, a linear interpolation between row effect and column effect and the overall total effect is:

$$\hat{\mu}(s) = \hat{a} + \hat{r}_k + \left(\frac{y - y_k}{y_{k+1} - y_k}\right)(\hat{r}_{k+1} - \hat{r}_k) + \hat{c}_l + \left(\frac{x - x_l}{x_{l+1} - x_l}\right)(\hat{c}_{l+1} - \hat{c}_l)$$

Where s=(x,y) in the region bounded by the four nodes $(x_l,y_k)$, $(x_{l+1},y_k)$, $(x_l,y_{k+1})$, $(x_{l+1},y_{k+1})$, where $x_l<x_{l+1}$ and $y_k<y_{k+1}$. For observations lie outside the grid an extrapolation technique used in MPK was constructed by the following formula. Suppose $x<x_1$ then,

$$\hat{\mu}(s) = \hat{a} + \hat{r}_k + \left(\frac{y - y_k}{y_{k+1} - y_k}\right)(\hat{r}_{k+1} - \hat{r}_k) + \hat{c}_1 + \left(\frac{x - x_1}{x_2 - x_1}\right)(\hat{c}_2 - \hat{c}_1)$$

For $y<y_1$, we may get

$$\hat{\mu}(s) = \hat{a} + \hat{r}_1 + \left(\frac{y - y_1}{y_2 - y_1}\right)(\hat{r}_2 - \hat{r}_1) + \hat{c}_l + \left(\frac{x - x_l}{x_{l+1} - x_l}\right)(\hat{c}_{l+1} - \hat{c}_l)$$

The median polish residuals R(.) can be considered to be stationary. Therefore, the residuals can be analyzed by using the ordinary Kriging, [2]. Hence,

$$\hat{R}(s) = \sum_{i=1}^{n} \lambda_i R(s_i)$$

According to equation (4) we may have,

$$\hat{Z}(s) = \hat{a} + \hat{r}_k + \left(\frac{y - y_k}{y_{k+1} - y_k}\right)(\hat{r}_{k+1} - \hat{r}_k) + \hat{c}_l + \left(\frac{x - x_l}{x_{l+1} - x_l}\right)(\hat{c}_{l+1} - \hat{c}_l) + \sum_{i=1}^{n} \lambda_i R(s_i)$$

Actually, $\hat{Z}(s)$ is an exact interpolator that uses a linear extrapolation method to extrapolate the points inside a high resolution grid which is finer than the original low resolution grid.

### 4. Biharmonic Spline Interpolation

The method of minimum curvature is an old and ever-popular approach for constructing smooth surfaces from irregularly spaced data. In one-dimensional case, the minimum curvature method leads to the natural cubic spline interpolation. In two-dimensional case, a surface can be interpolated with biharmonic spline. A simpler algorithm for finding the minimum curvature surface that passes through a set of nonuniformly spaced data points, [8].

Obviously, the spline has zero fourth derivative, hence; the spline will satisfy the biharmonic equation as:

$$\frac{d^4\phi}{dx^4} = 6\delta(x) \qquad (5)$$

The particular solution to (5) is

$$\phi(x) = |x|^3$$

When this green function is used to interpolate N data points, $w_i$, located at xi the problem is

$$\frac{d^4 w}{dx^4} = \sum_{j=1}^{N} 6\alpha_j \delta(x - x_j) \qquad (6)$$

$$w(x_i) = w_i \qquad (7)$$

The particular solution to equations (6) and (7) is a linear combination of points forced Green functions centered at each data point. Therefore, we have

$$w(x) = \sum_{j=1}^{N} \alpha_j |x - x_j|^3 \qquad (8)$$

The strength of each point force, $\alpha_j$, is found by solving linear system

$$w_i = \sum_{j=1}^{N} \alpha_j |x - x_j|^3$$

If slopes, $S_i$, are used rather than values, then the $\alpha_j$'s are determined by solving the following linear system

$$S_i = 3\sum_{j=1}^{N} \alpha_j |x_i - x_j|(x_i - x_j)$$

Once $\alpha_j$'s are determined, the biharmonic function w(x) can be evaluated at any point using equation (8), [8].

### 5. Improved median polish kriging

The method of Improved Median Polish Kriging (IMPK) is an improvement of the traditional Median Polish Kriging. The improvement is concerned with the estimation of the mean function μ(.), i.e. the Biharmonic spline interpolation is replaced by the linear spline interpolation. Kriging and spline are formally alike, but practically different. Both disciplines can benefit from each other's knowledge base. There is a formal connection between these two very important methods of interpolation, but there is a large divergence in how they are applied and how their results are interpreted. In this article, we must mention the method of modified median polish Kriging (MMPK) proposed by [1] which uses a different technique to estimate the mean function μ(.). To distinguish between our proposed method (IMPK) and MMPK, the method of MMPK using universal Kriging to estimate the mean function μ(.) which needs more in computer time. Since each unobserved point in the high resolution grid for mean function μ(.) needs to be estimated using universal Kriging, what about time consuming here?. Alternatively, IMPK uses Biharmonic interpolation to interpolate and extrapolate the nodes inside and outside the high resolution grid. Now, we will derive the general formula for the IMPK using Biharmonic spline interpolation.

For N data in two dimensions the problem is:

$$\nabla^4 w(s) = \sum_{j=1}^{N} \alpha_j \delta(s - s_j)$$

Where $s \in \hat{u}(s)$ and $\nabla^4$ is the biharmonic operator and s is the unobserved data point in m-dimension. Then, the general solution is

$$w(s) = \sum_{j=1}^{N} \alpha_j \phi_m (s - s_j) \qquad (9)$$

Where $\phi_m$ can be found for each dimension in Table(1).

Table 1. Biharmonic Green Functions [8]

| Number of dimensions (m) | Green function $\phi_m$ |
|---|---|
| 1 | \|x\|3 |
| 2 | \|x\|2(ln\|x\|-1) |
| 3 | \|x\| |
| 4 | ln\|x\| |
| 5 | \|x\|-1 |
| 6 | \|x\|-2 |

| m | \|x\|4-m |
|---|---|

Since, our concern in this work is in two-dimensional simulation space, then $\phi_2$ can be substituted in (9) leading to:

$$w(s) = \sum_{j=1}^{N} \alpha_j \left( |s - s_j| 2\ln(|s - s_j|) - 1 \right) \quad (10)$$

Which is the Biharmonic interpolator of the mean function µ(.) in a high resolution grid. Therefore, µ(s) can be substituted by w(s) in equation (4), yields

Now, our concern is to estimate the residuals R(s) in equation (4). Basically, the same technique used by the original MPK. The residual values {R($s_i$) : 1,2 …,n} can be used as a new data set to allow new fresh observations as low resolution grid that can be used by ordinary Kriging to predict all the residual values on a finer grid. Therefore, we can predict the values of the residuals as:

$$\hat{R}(s) = \sum_{i=1}^{N} \lambda_i R(s_i)$$

Hence, equation (4) can be written as:

$$\hat{Z}(s) = \hat{u}(s) + \hat{R}(s)$$

Therefore, w(s) is an exact approximation of $\hat{u}(s)$. Then, we can write (4) in an approximated (predicted) form as:

$$\hat{Z}(s) = w(s) + \hat{R}(s)$$

And the final formula for IMPK formula is

$$\hat{Z}(s) = \sum_{j=1}^{N} \alpha_j \left( |s - s_j| 2\ln(|s - s_j|) - 1 \right) + \sum_{i=1}^{N} \lambda_i R(s_i)$$

The final formula must be smoother because it is non-linear function of the unobserved data points. Conversely, the traditional MPK uses linear interpolation which is linear, i.e. has first order approximation $O(1)$.

## 6. Experimental Results

Experimentally, we have used the standard coal-Ash data given by [2], which is standard data set to investigate the prediction of Kriging model. This data collection has been used by many authors as a standard two-dimensional input data for the response surfaces (metamodels). In this article, we have showed that the newly developed methods did not only prove to be of academic interest, but also very useful in simulating two-dimensional surfaces.

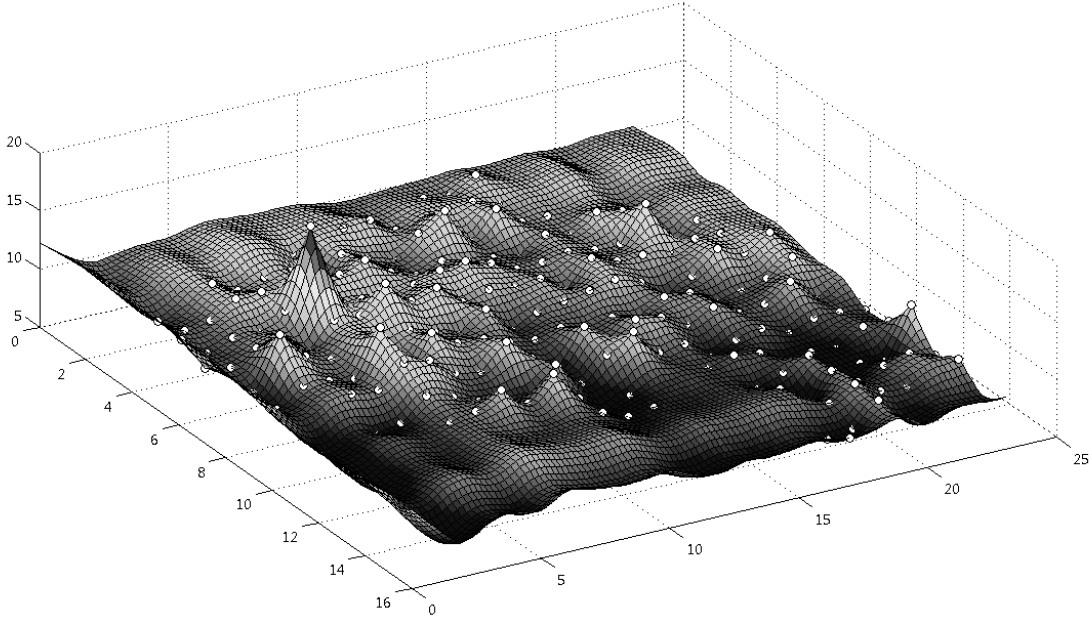

(a)

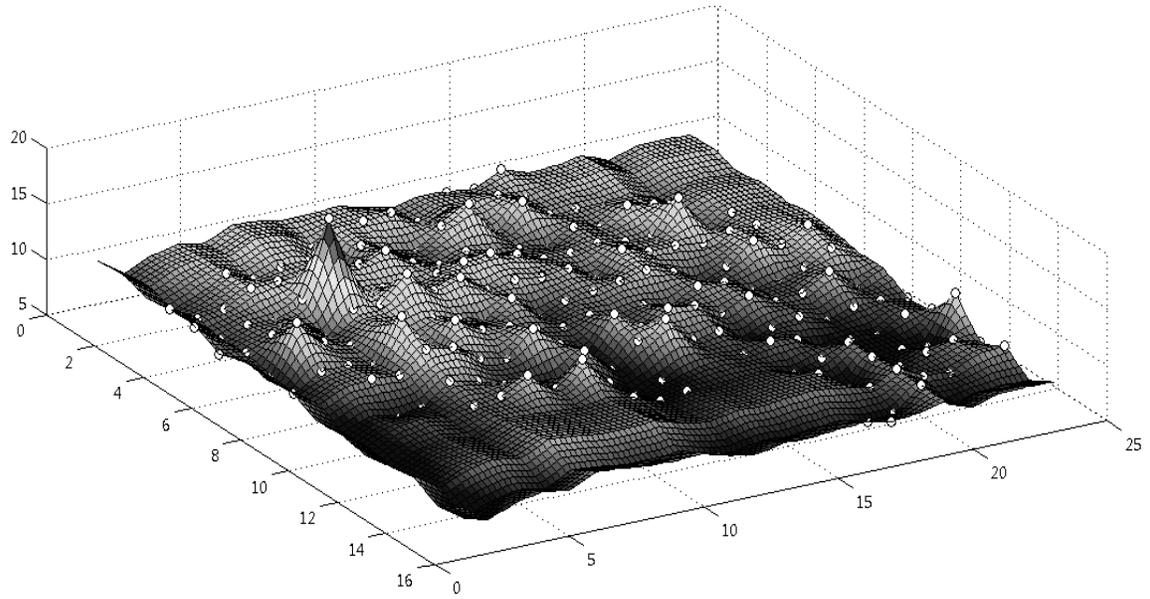

(b)

Fig. 1: Surface prediction using (a) IMPK (b) MPK

Graphically speaking, we can see that figure (1.a) have smoother surface than figure (1.b). Also, the cross validation method has been used here to validate our Improved Median Polish Kriging (IMPK). The RMSE used to validate IMPK is given in table (2) as:

Table 2: Root Mean Square Error for IMPK and MPK

|  | IMPK | MPK |
|---|---|---|
| RMSE | 1.170527 | 1.701783 |

According to figure (2), we can see that the variance in the IMPK method is smoother than MPK. Hence, the prediction may be more adequate in the increasing and decreasing sub-surface in the original surface.

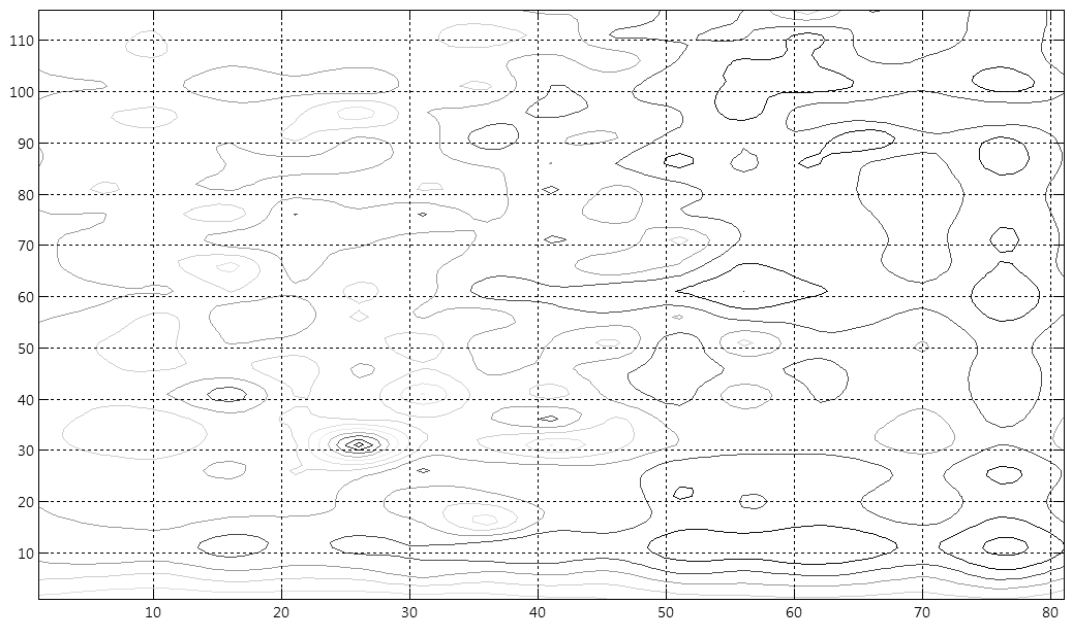

(a)

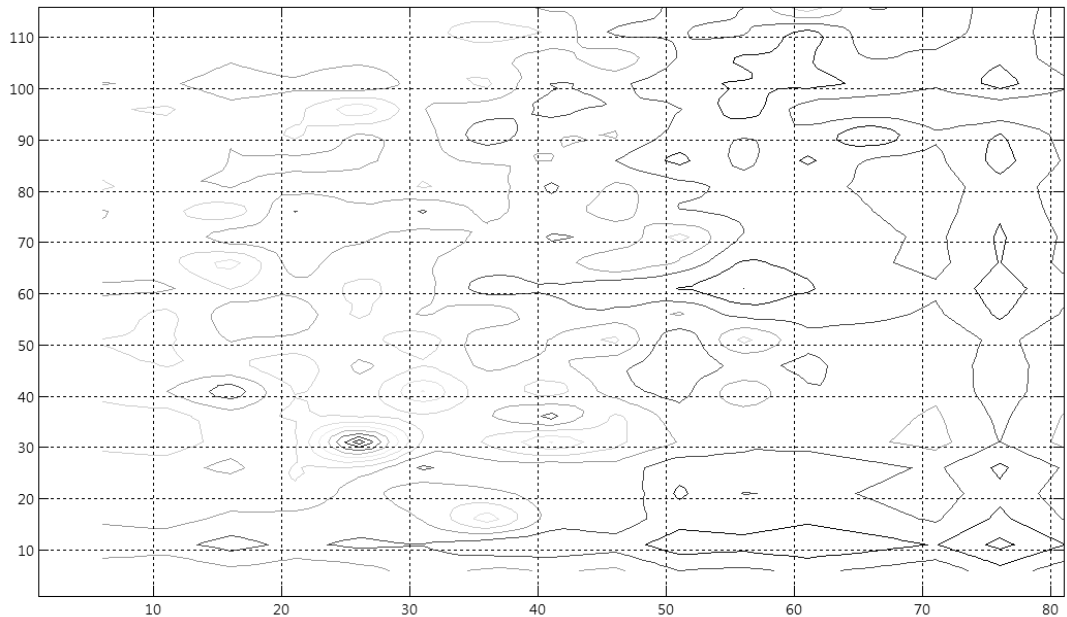

Fig. 2: Error maps for (a) IMPK (b) MPK

## 7. Conclusions

In this article, we developed a novel method to predict two-dimensional surface in any simulated metamodel. The interpolation type arises as a critical point in the prediction. The resulted surface gave a smoother shape than traditional MPK that will be suitable to mimics the original surface (system). The pros and cons of the new method have been presented. The first recommendation for the future work is to generalize this method in three-dimensions and multi-dimensions. The second recommendation is the use of Hermit interpolation in the finer grid interpolation since its structure could be suitable in median Polish Kriging because it is a robust non-linear interpolation.

**References**

[1] Berke O., 2001. Modified Median Polish Kriging and its application to the Wolfcamp-Aquifer data. Environmentrics, 12: 731-748.

[2] Cressie, N., 1993. Statistics for spatial data. John Wiley & Sons, Inc., New York.

[3] Gunes H., Sirisup S., Karniadakis G E, Gappy data: To Krig or not to Krig?, 2006. Journal of Computational Physics 212, 358–382.

[4] Kleijnen, J. P. C., 2007. Kriging Metamodeling in Simulation: A Review, ISSN 0924-7815 Tilburg University.

[5] Matheron G., 1967. Kriging or Polynomial Interpolation Procedures?, mathematical geology transactions volume LXX, pp 240-244.

[6] Sacks J., Welch W.J., Mitchell T.J., Wynn H.P. 1989. Design and Analysis of Computer Experiments, Statistical Science, 4, pp. 409-435.

[7] Sandor B., Lambert M., Vazquez E. and Gyimothy S., 2010. Combination of Maximin and Kriging Prediction Methods for Eddy-Current Testing Database Generation, Journal of Physics: Conference Series 255 (2010) 012003.

[8] Sandwell, D.T., 1987. Biharmonic Spline Interpolation of GEOS-3 and SEASAT Altimeter data. Geophysical Research Letters, 2, 139-142.

[9] Van Beers, W. C. M. 2005. Kriging Metamodeling in Discrete-Event simulation An overview, Proceedings of the 2005 Winter Simulation Conference.

[10] Van Beers, W. C. M. and Kleijnen J. P. C.. 2004. Kriging interpolation in simulation: a survey, Proceedings of the 2004 Winter Simulation Conference.

[11] Van Beers, W. C. M. and. Kleijnen J. P. C. 2003. Kriging for interpolation in random simulation, Journal of the Operational Research Society, 54: 255-262.